\documentclass[letter,scriptaddress,twocolumn]{revtex4}

	\usepackage{amsmath}
	\usepackage{amssymb} 
	\usepackage[sort&compress]{natbib}
	\usepackage{makeidx}
	\usepackage{amsfonts}
	\usepackage[ansinew]{inputenc}
	\usepackage[usenames,dvipsnames]{pstricks}
	\usepackage{epsfig}
	\usepackage{pst-grad} 
	\usepackage{pst-plot} 
	\usepackage[colorlinks,hyperindex]{hyperref}
	\hypersetup
	{
		colorlinks,%
		citecolor=blue,%
		linkcolor=blue,%
		urlcolor=black,%
	}

\newcommand{\beqn}{\begin{equation}}
\newcommand{\eeqn}{\end{equation}}
\newcommand{\beqna}{\begin{eqnarray}}
\newcommand{\eeqna}{\end{eqnarray}}
\makeindex
\begin{document}
\title{Radial flow in a multi phase transport model at FAIR energies}
\author{Soumya Sarkar$^{1,2}$, Provash Mali$^1$, Somnath Ghosh$^1$ and Amitabha Mukhopadhyay$^1$}
\email{amitabha$_$62@rediffmail.com}
\affiliation{$^1$Department of Physics, University of North Bengal, Siliguri 734013,  India}
\affiliation{$^2$Department of Physics, Siliguri College, Siliguri 734001, India}
\begin{abstract}
Azimuthal distributions of radial velocities of charged hadrons produced in nucleus-nucleus $(AB)$ collisions are compared with the corresponding azimuthal distribution of charged hadron multiplicity in the framework of a multiphase transport (AMPT) model at two different collision energies. The mean radial velocity seems to be a good probe for studying radial expansion. While the anisotropic part of the distributions indicates a kind of collective nature of radial expansion, the isotropic part characterizes a thermal motion. The present investigation is carried out keeping the upcoming Compressed Baryonic Matter (CBM) experiment to be held at the Facility for Anti-proton Ion Research (FAIR) in mind. As far as high-energy heavy-ion interactions are concerned, CBM will supplement the Relativistic Heavy Ion Collider (RHIC) and Large Hadron Collider (LHC) experiments. In this context our simulation results at high baryochemical potential would be interesting, when scrutinized from the perspective of almost a baryon free environment achieved at RHIC and LHC.
\end{abstract}
\maketitle
\section{Introduction}
When two heavy nuclei collide with each other at high-energy it is expected that a color deconfined state composed of strongly coupled quarks and gluons is formed. The properties of such a state, formally known as Quark-Gluon Plasma (QGP) \cite{Nagamiya84}, are governed by the rules of quantum chromodynamics (QCD). In order to understand the bulk properties of this extended QCD state and to understand the dynamical processes that might be involved in its formation and subsequent decay, over last three decades or so QGP has been widely searched for in many high-energy experiments \cite{Karsch05}. Of all the efforts in this regard, the study of the final state particles with respect to the reaction plane of an $AB$ collision, spanned by the beam direction and the impact parameter vector, has been a popular technique that can characterize the thermodynamic and hydrodynamic properties of the QGP matter \cite{Ollitrault92, Voloshin96}. The Fourier decomposition of the aforementioned distributions has been widely employed to explore the collective behavior of the final state particles. More specifically, the second harmonic coefficient, traditionally known as the elliptic flow parameter $(v_2)$ is of special interest \cite{Voloshin08}. The $v_2$ results obtained from the RHIC and LHC experiments show considerable hydrodynamical behavior of the matter present in the overlapping zone of the colliding nuclei, an intermediate ``fireball'' that gets thermalized within a very short time interval ($< 1$ fm/c), and  subsequently expands almost like a ``perfect fluid" having a very small ratio of shear viscosity to entropy density \cite{Sorge97,Huovinen01,Luzum08}. In RHIC \cite{Adler01,Abelev07,Adamczyk13,Adare07,Back05} and LHC \cite{Chatrchyan12,Aad12, Aamodt11, Abelev15} experiments the $v_2$ parameter has been widely studied as a function of the centrality of the collision, transverse momentum $(p_{_T})$ and rapidity $(y)$ or pseudorapidity $(\eta)$ of produced particles, for different colliding systems and at varying collision energies. Using the AMPT model, presence of anisotropy in the azimuthal distribution of transverse rapidity $(y_{_T})$ has been investigated in Au+Au collision at $\sqrt{s_{_{NN}}} = 200$ GeV \cite{Li13}. Recently, we have reported some simulation results on the anisotropy present in the azimuthal distribution of $p_{_T}$ of the emitted charged hadrons that has relevance to the radial flow of charged hadrons produced at FAIR energies \cite{Sarkar17a}. The CBM experiment is dedicated to explore the deconfined QCD matter at high baryon density and low to moderate temperature. It is a fixed target experiment being designed with incident energy range $E_{\rm lab}= 10$ - $40$ GeV per nucleon, which is expected to produce partonic matter of density $6$ to $12 $ times the normal nuclear matter density at the central rapidity region \cite{Stocker86}. But at the same time it should be kept in mind that our present understanding of collective flow of hadronic/partonic matter at this energy region is constrained by the availability of only a few experimental results \cite{NA49}. Therefore, to get any idea about the expected behavior of any variable or parameter that is relevant in this regard, we have to rely mostly upon event generators and models. In this article we present some basic simulated results on the (an)isotropy of the radial velocity of charged hadrons produced in Au+Au collisions  at $E_{\rm lab} = 10A$ and $40A$ GeV using the AMPT model \cite{AMPTa, AMPTb}. The paper is organized as follows, a brief description of the methodology used in this analysis is given in Section\,\ref{meth}, the AMPT model is summarily described in Section\,\ref{ampt}, our simulation based results are discussed in Section\,\ref{results}, and finally in Section\,\ref{conclusion} our observations are listed.
\section{Methodology}
\label{meth}
Before the collisions, the nucleons belonging to individual nucleus possess only longitudinal degrees of freedom. Transverse degrees of freedom are excited into them only after an interaction takes place. In mid-central collisions the overlapping area of the colliding nuclei is almond shaped in the transverse plane. This initial asymmetry in the geometrical shape gives rise to different pressure gradients along the long and short axis of the overlapping zone, and correspondingly to a momentum space asymmetry in the final state. As a result, if the matter present in the intermediate ``fireball'' exhibits a fluid like behavior, then a collective flow of final state particles is observed, which is reflected in the azimuthal distribution of particle number as well as in the azimuthal distribution of an appropriate kinematic variable like $p_{_T}$, $y_{_T}$ and the transverse or radial velocity $v_{_T}$ \cite{Li12}. The radial velocity has two components, the radial flow velocity and the velocity due to random thermal motion. For an ideal fluid the radial flow velocity should be isotropic. However, for a non-ideal viscous fluid, as the case may be at FAIR conditions, the shear tension is proportional to the gradient of radial velocity along the azimuthal direction, which again is related to the anisotropy of radial velocity \cite{Li13}. We introduce the transverse (radial) velocity as 
\beqn
v_{_T} = \frac{p_{_T}}{E} = \frac{p_{_T}}{m_{_T} \cosh y},
\eeqn
where $E=m_{_T} \cosh y$ is the energy of the particle, $m_{_T} = \sqrt{m_{_0}^2 + p_{_T}^2}$ is its transverse mass, $m_{_0}$ is the particle rest mass, and $y$ is its rapidity. For a large sample of events the total radial velocity $\langle V_{_T}(\phi_m) \rangle$ of all particles falling within the $m$-th azimuthal bin is defined as,
\begin{equation}
	\langle V_{_T}(\phi_m) \rangle = \frac{1}{N_{ev}} \sum_{j=1}^{N_{ev}}\,\sum_{i=1}^{n_{_m}} v_{_{T,i}}(\phi_{_m})
	\label{tot-vel}
\end{equation}
where $v_{_{T,i}}(\phi_m)$ is the radial velocity of the $i$-th particle, $n_{_m}$ is the total number of particles present in the $m$-th bin, $N_{ev}$ is the number of events under consideration and $ \langle~\rangle $ denotes an averaging over events. In this paper we have chosen the transverse velocity as the basic variable in terms of which the azimuthal asymmetry has been studied and compared the results obtained thereof with those of the azimuthal asymmetry associated with the charged particle multiplicity distribution. An azimuthal distribution of $\langle V_{_T}(\phi_m) \rangle$ contains information of the multiplicity as well as that of the radial expansion. By taking an average over particle number the mean transverse velocity $\langle \langle v_{_T}(\phi_m) \rangle \rangle$ is introduced as
\begin{equation}
	\langle \langle v_{_T}(\phi_m) \rangle \rangle = \frac{1}{N_{ev}}\sum_{j=1}^{N_{ev}} \frac{1}{N_m}\sum_{i=1}^{N_m} v_{_{T,i}}(\phi_m)
	\label{mean-vel}
\end{equation}	
where $\langle \langle ~ \rangle \rangle$ represents first an average over all particles present in the $m$-th azimuthal bin and then over all events present in the sample. This double averaging reduces the multiplicity influences significantly, and the corresponding distribution measures only the radial expansion. In this context we must mention that the mean radial velocity actually consists of contributions coming from three different sources, the average isotropic radial velocity, the average anisotropic radial velocity, and the average velocity associated with thermal motion. It should be noted that both radial and thermal motion contribute to the isotropic velocity of the distribution. Like the azimuthal distribution of charged particle multiplicity $d\left<N_{\rm ch}\right>/d\phi$, it is also possible to expand the azimuthal distributions of total and mean transverse velocities in Fourier series as, 
\begin{eqnarray}
\frac{d\left<V_{_T}\right>}{d\phi} &\approx & v_{_0}\left(\left<V_{_T}\right>\right)\left[1 + 2\,v_{_2}\left(\left<V_{_T}\right>\right)\cos(2\phi)\right], \label{VT-vTx} \\
\frac{d\left<\left<v_{_T}\right>\right>}{d\phi} &\approx & v_{_0}\left(\left<\left<v_{_T}\right>\right>\right)[1 + 2\,v_{_2}\left(\left<\left<v_{_T}\right>\right>\right)\cos(2\phi)].
\label{VT-vT}
\end{eqnarray}
In these expansions only the leading order terms ($n = 0~\mbox{and}~ 2$) are retained. The anisotropy present in any of the distributions [Equations.\,(\ref{VT-vTx}) and (\ref{VT-vT})] is quantified by the second Fourier coefficient $v_2$, whereas $v_0$ is a measurement of the isotropic flow.
\section{AMPT Model}
\label{ampt}
Transport models are best suited to study $AB$ collisions at the energy range under our consideration. Since transport models treat chemical and thermal freeze-out dynamically, they have the ability to describe the space-time evolution of the hot and dense ``fireballs'' created in collision between two heavy nuclei at relativistic energy. As mentioned before, in this simulation study we use the AMPT model with partonic degrees of freedom, the so-called string melting version, with the expectation that under FAIR conditions a transition from the QGP state to the hadronic state may take place at high baryon density and low to moderate temperature. Previous calculations have shown that flow parameters consistent with experiment, can be developed through AMPT, and the model can successfully describe different aspects of the collective behavior of hadronic/partonic matter produced in $AB$ interactions \cite{AMPT1, AMPT2, Partha10, Nasim10}. The string melting version of AMPT should be even more appropriate to model particle emission data, where a transition from nuclear matter to deconfined QCD state is expected. AMPT is a hybrid model where the primary particle distribution and other initial conditions are taken from the heavy-ion jet interaction generator (HIJING) \cite{HIJING}, and Zhang's parton cascade (ZPC) formalism \cite{ZPC} is used in subsequent stages. Note that the ZPC model includes only parton-parton elastic scattering with an in-medium cross section derived from pQCD, the effective gluon screening mass being taken as an adjustable parameter. In the string melting version of the AMPT model all hadrons are produced from string fragmentation like that in the HIJING model. The strings are converted into valence quarks and antiquarks, are allowed to interact through the ZPC formalism, and propagate according to a relativistic transport model \cite{AMPTb}. Finally, the quarks and antiquarks are converted to hadrons via a quark coalescence formalism.
\section{Results and Discussion}
\label{results}
\begin{figure}[t]
\centering
\vspace{-1cm}
\includegraphics[width=0.45\textwidth]{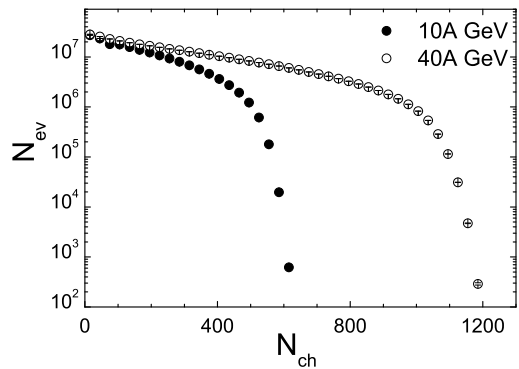}
\vspace{-.5cm}
\caption{Charged hadron multiplicity distribution in Au+Au collisions at $E_{\rm lab} = 10A$ and $40A$ GeV. }
\vspace{-0.2cm}
\label{multi}
\end{figure}

\begin{figure}[t]
\centering
\vspace{-0.5cm}
\includegraphics[width=0.45\textwidth]{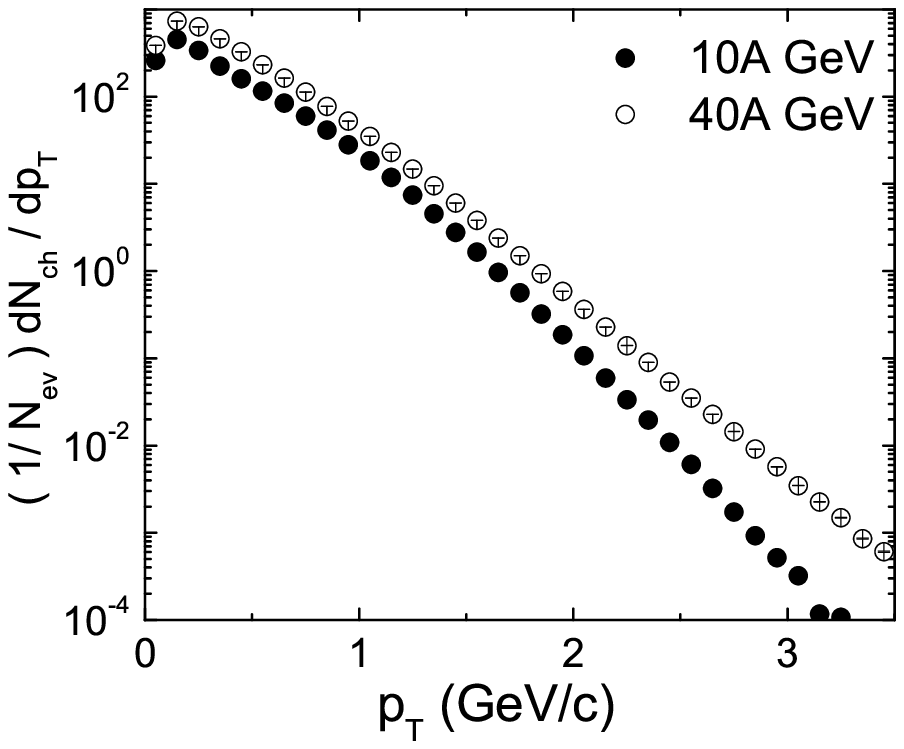}
\vspace{-.5cm}
\caption{Charged hadron $p_{_T}$ distribution in Au+Au collisions at $E_{\rm lab} = 10A$ and $40A$ GeV.}
\vspace{-0.2cm}
\label{pt}
\end{figure}

\begin{figure}[tbh]
\centering
\vspace{-1cm}
\includegraphics[width=0.45\textwidth]{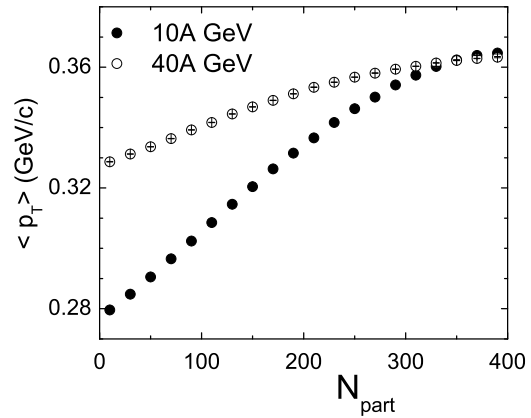}
\vspace{-.5cm}
\caption{Average transverse momentum of charged hadrons as a function of $N_{\rm part}$ in Au+Au collisions at $E_{\rm lab} = 10A$ and $40A$ GeV.}
\vspace{-0.2cm}
\label{npart-pt}
\end{figure}

\begin{figure}[tbh]
\includegraphics[width=0.45\textwidth]{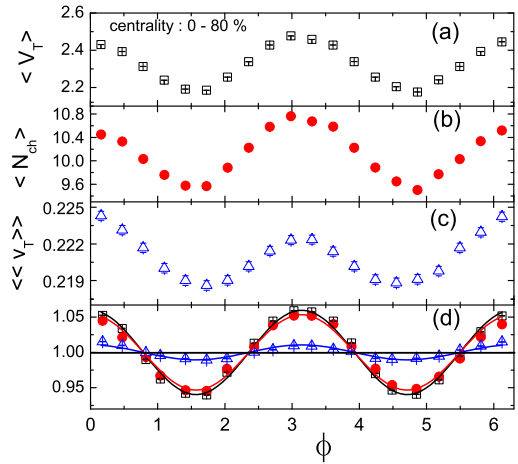}
\caption{(Color online) Azimuthal distribution of (a) total radial velocity, (b) multiplicity, (c) mean radial velocity and (d) all the aforesaid quantities properly normalized for charged hadrons produced in Au+Au collisions at $E_{\rm lab} = 40A$ GeV.}
\vspace{-0.5cm}
\label{azimuth}
\end{figure}

In this section we describe our results obtained from Au+Au minimum bias events simulated by the AMPT model (string melting version) at $E_{\rm lab} = 10A$ and $40A$ GeV. A representative value of the parton scattering cross section ($\sigma=3$ mb) is used in this analysis. The $\sigma$ value is so chosen as to match with a previously studied collective behavior at FAIR energies \cite{Partha10}. We have indeed compared the NA49 results \cite{NA49} on the $p_{_T}$ dependence of elliptic flow parameter $v_2$ by varying $\sigma$ over a range of $0.1$ to $6$ mb. We have seen that even though the $\sigma$ values differ almost by an order of two, the corresponding differences in the simulated $v_2$ values are not that significant \cite {Sarkar17b}. The size of each sample of Au+Au events used in this analysis is one million. We begin with the multiplicity distribution of charged hadrons, represented schematically in Figure\,\ref{multi}. The nature of the distributions is more or less similar at both energies concerned. However, the average and the highest multiplicities are naturally quite larger at $E_{\rm lab} = 40A$ GeV. In Figure\,\ref{pt} we plot the $p_{_T}$ distributions of charged hadrons. As expected with increasing $p_{_T}$ we observe an approximately exponential fall in the particle number density. It is interesting to note that at low $p_{_T}$ values, up to $p_{_T}$ $\approx 1.5$ GeV/c, the slopes of the distributions at both energies hardly differ, but at high $p_{_T}$ beyond $p_{_T}$ $= 2.0$ GeV/c the slope values are considerably different, this being stiffer at $E_{\rm lab} = 10A$ GeV. The inverse slope can be related to the temperature of the intermediate ``fireball'' as and when it achieves thermal equilibrium. Therefore, with varying collision energies, while the particles produced in soft processes correspond to almost same source temperature, those produced in hard processes belong to a higher source temperature for a higher collision energy. A Monte Carlo Glauber (MCG) model \cite{Miller07} is employed to characterize the geometry of an $AB$ collision. Using the MCG model the average transverse momentum $\langle p_{_T}\rangle$ of particles produced in $AB$ collisions belonging to a particular centrality can be determined. In Figure\,\ref{npart-pt} such a plot of $\langle p_{_T}\rangle$ against the number of participating nucleons $(N_{\rm part})$, a measure of the centrality of the collision, is graphically shown. At the two collision energies considered at low $N_{\rm part}$ the $\langle p_{_T}\rangle$ values are significantly different, $\langle p_{_T}\rangle$ increases almost linearly with increasing centrality, and each distribution saturates at similar value, $\langle p_{_T}\rangle \approx 0.365$ GeV/c. The fact that at the highest centrality the saturation value of $\left<p_{_T}\right>$ of produced particles is almost independent of the incident beam energy is perhaps due to kinematic reasons. The transverse degree of freedom that was absent before the collision took place, is excited into the interacting $AB$ system due to multiple nucleon-nucleon $(NN)$ scattering and re-scattering. Our results indicate that the degree of such excitations predominantly depends on the number of binary collisions, which should be same for the most central collisions of the same colliding system (Au + Au) at the two collision energies under consideration.
In Figure\,\ref{azimuth} we present the azimuthal distributions of (a) the total radial velocity $\langle V_{_T} \rangle$, (b) the multiplicity $(N_{\rm ch})$ and (c) the mean radial velocity $\langle\langle v_{_T} \rangle\rangle$ of charged hadrons produced at $E_{\rm lab} = 40A$ GeV in the mid-rapidity region, $\Delta y = \pm 1.0$ symmetric about the central value $y_0$, within the $0-80\%$ centrality range. Presence of anisotropy in all three distributions is clearly visible. It is also observed that while all three distributions exhibit same periodicity, their amplitudes are quite different. In order to show that all three distributions can analytically be described by a single function like $N\left[1+\alpha\,\cos (2\phi)\right]$ without significant contributions coming from other harmonics, we fit the distributions with exactly the same relative vertical axis range with respect to the value of the parameter $\alpha$ centred around the same value of the other parameter $N$ (here $N=1.0$), and plot them together in  Figure\,\ref{azimuth}(d) along with the respective fitted lines. When appropriately scaled, we find that the elliptic anisotropy in the distribution of total radial velocity is almost equal in magnitude as that coming from the anisotropy in multiplicity distribution. In comparison, corresponding anisotropy in the mean radial velocity is quite small. The results at $E_{\rm lab} = 10A$ and $40A$ GeV are qualitatively similar.
\subsection{Centrality dependence of $v_{_2}$ and $v_{_0}$ }
\begin{figure}[tbh]
\centering
\vspace{-0.7cm}
\includegraphics[width=0.5\textwidth]{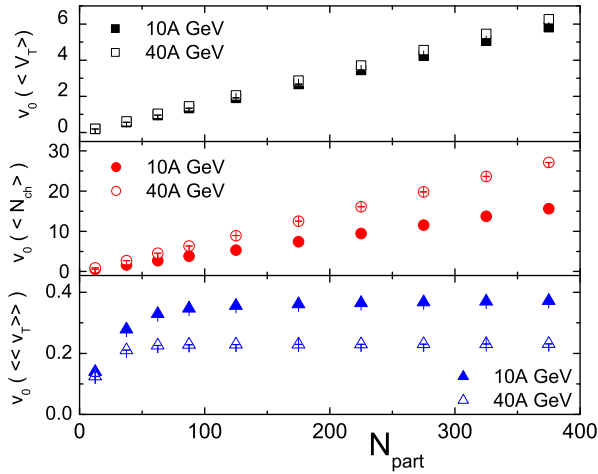}
\vspace{-.5cm}
\caption{(Color online) Centrality dependence of isotropic flow coefficient $v_0$ obtained from the azimuthal distributions of total radial velocity, multiplicity and mean radial velocity in Au+Au collision at $E_{\rm lab} = 10A$ and $40A$ GeV.}
\vspace{-0.2cm}
\label{v0_npart}
\end{figure}

Elliptic flow results from interactions between particles comprising the intermediate ``fireball'', and hence it is an useful probe for the identification of local thermodynamic equilibrium. The $v_2$ values are smaller for the extreme central and peripheral collisions, which can be explained in terms of the initial geometric effects and the pressure gradient produced thereof \cite{Alver07}. In the hydrodynamical limit $v_2$ is proportional to the elliptic eccentricity $(\varepsilon_2)$ of the overlapping region of the colliding nuclei, whereas in the low density limit $v_2/\varepsilon_2$ is proportional to the product of the rapidity density of charged particles $d\left<N_{\rm ch}\right>/dy$ and inverse of the overlapping area of the colliding nuclei. It is believed that the centrality dependence of elliptic flow provides valuable information regarding the degree of equilibration achieved by the intermediate ``fireball'', and also regarding the characteristics of (re)scattering effects present therein \cite{Voloshin00}. Some model based results at FAIR energies can be found in \cite{Partha10,Sarkar17a,Sarkar17b}.\\
\begin{figure}[tbh]
\centering
\vspace{-1cm}
\includegraphics[width=0.5\textwidth]{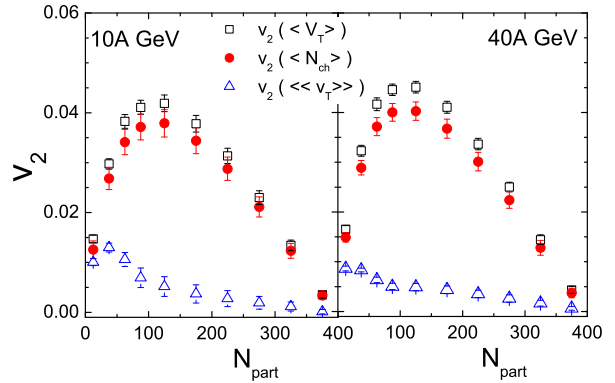}
\vspace{-.5cm}
\caption{(Color online) Centrality dependence of anisotropy parameter $v_2$ obtained from the azimuthal distributions of total radial velocity, multiplicity, and mean radial velocity in Au+Au collision at $E_{\rm lab} = 10A$ and $40A$ GeV.}
\vspace{-0.2cm}
\label{v2-npart}
\end{figure}
\\
In Figure\,\ref{v2-npart} we compare the centrality dependence of the $v_2$ parameter obtained from distributions of all three variables under consideration. The overall centrality dependence is found to be similar for $v_2(\left<V_{_T}\right>$ and $v_2(\left<N_{\rm ch}\right>$. However, the $v_2(\left<\left< v_{_T}\right>\right>$ which are quite small in comparison with the $v_2$ values obtained from the other two variables, behave quite differently. All three variations however, are consistent with our observations of Figure\,\ref{azimuth}(d). It is to be noted that the anisotropy in mean radial velocity, which describes the radial expansion, is significantly smaller than that of the corresponding multiplicity distribution in the mid-central region. In this regard we also intend to scrutinize the effects of the collision energy involved. It is observed that $v_2(\langle V_{_T} \rangle)$ and $v_2(\langle N_{\rm ch} \rangle)$ at $E_{\rm lab} = 40A $ GeV are marginally higher than those at $10A$ GeV, a general feature of any $v_2$ result, which has been confirmed over a wide energy range. The $v_2(\langle \langle v_{_T} \rangle \rangle)$ values are not significantly different at the two collision energies involved. We expect that the isotropy parameter $(v_0)$ of all aforesaid distributions are also of certain importance and we graphically plot the results in Figure\,\ref{v0_npart}. The $v_0$ values associated with $\langle V_{_T}\rangle$ and $\langle N_{\rm ch}\rangle$ distributions show a linear dependence with increasing $N_{\rm part}$, being highest in the most central events. This feature of $v_0$ can be ascribed to the fact that the azimuthally integrated magnitude of transverse flow increases with increasing centrality of the collisions. On the contrary, an increasing trend in the $v_0(\langle \langle v_{_T} \rangle \rangle)$ values with increasing $N_{\rm part}$ is restricted only to the  peripheral collisions, and beyond $N_{\rm part}\approx 80$ the $v_0(\langle \langle v_{_T} \rangle \rangle)$ values achieve a saturation, being nearly independent of the centrality of the collisions. A significant energy dependence of $v_0$ is also observed for all the variables considered in this analysis. We do not see any significant energy dependence in the variation of $v_0(\langle \langle V_{_T} \rangle \rangle)$ with $N_{\rm part}$. The $v_0(\langle N_{\rm ch}\rangle)$ values are however consistently higher at $E_{\rm lab}=40A$ GeV than those at $E_{\rm lab}=10A$ GeV, the difference getting larger with increasing $N_{\rm part}$. Once again $v_0(\langle \langle v_{_T} \rangle \rangle)$ behaves quite differently in this regard. The values at $E_{\rm lab} = 10A $ GeV are consistently higher than those at $40A$ GeV. We may recall that the mean radial velocity has been defined in a way such that the multiplicity effects are removed. Therefore, we conclude that the particle multiplicity plays a dominant role to determine the total transverse flow, and a higher energy input results in a lower amount of azimuthally integrated transverse flow. 
\subsection{Transverse momentum dependence of $v_2$ and $v_0$}
\begin{figure}[tbh]
\centering
\vspace{-.8cm}
\includegraphics[width=0.5\textwidth]{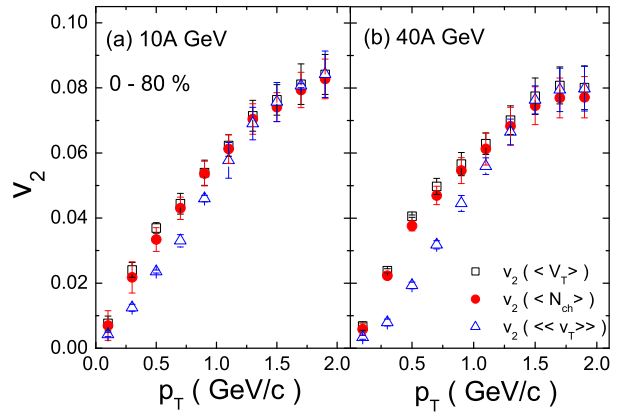}
\vspace{-.5cm}
\caption{(Color online) Transverse momentum dependence of anisotropy parameter $v_2$ obtained from the azimuthal distributions of total radial velocity, multiplicity and mean radial velocity in Au+Au collision at $E_{\rm lab} = 10A$ and $40A$ GeV.}
\label{v2-pt}
\end{figure}

It is well known that the anisotropy coefficient $v_2$ depends on the $p_{_T}$ of charged hadrons. Hydrodynamics as well as resonance decays are expected to dominate at low $p_{_T}$, whereas high $p_{_T}$ particles are expected to stem out from the fragmentation of jets modified in the hot and dense medium of the intermediate ``fireball'' \cite{Aad12}. At FAIR energies the production of high $p_{_T}$ hadrons would be rare, and owing to statistical reasons we restrict our analysis up to $p_{_T} = 2.0$ GeV/c. $v_2$ arising from multiplicity distributions of the produced hadrons has been studied widely as a function of $p_{_T}$ using the data available from the experiments held at RHIC \cite{Adcox05} and LHC \cite{Aamodt10}. Simulation results under FAIR-CBM conditions utilizing the UrQMD, AMPT (default) and AMPT (string melting) models can be found in \cite{Sarkar17a,Partha10}. Figure\,\ref{v2-pt} depicts that the anisotropies present in $\langle N_{\rm ch} \rangle$, $\langle V_{_T} \rangle$ and $\langle \langle v_{_T} \rangle \rangle$ rise monotonically with increasing $p_{_T}$. At $E_{\rm lab}=40A$ GeV beyond $p_{_T}=1.5$ GeV/c there is a trend of saturation in the $v_2$ values extracted from all three variables. Once again we conclude that the multiplicity dominates over the radial velocity at a particular $p_{_T}$ bin, and $v_2(\langle N_{\rm ch} \rangle)$ and $v_2(\langle V_{_T} \rangle)$ are found to be almost equal in the $0 \leqslant p_{_T} \leqslant 2.0$ GeV/c range. Once we get rid of the multiplicity effects, the actual anisotropy present in the radial velocity comes out, which we can see in the plot of $v_2(\langle\langle v_{_T}\rangle\rangle)$ against $p_{_T}$ shown in the same diagram. As a result the $v_2(\langle\langle v_{_T}\rangle\rangle)$ are slightly different (lower) within $0.25 \leq p_{_T}\leq 1.25$ GeV/c. At FAIR energies, however, we do not find any noticeable deviation in the trend of this $p_{_T}$ dependence of $v_2$ from its nature observed at RHIC energies \cite{Li12}. 
\begin{figure}[tbh]
\centering
\vspace{-1cm}
\includegraphics[width=0.5\textwidth]{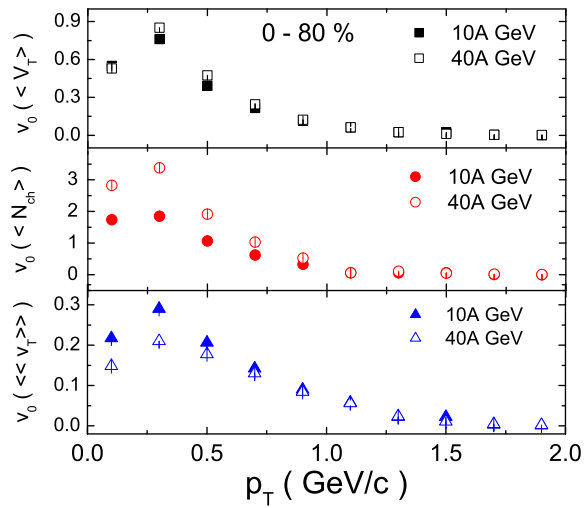}
\vspace{-.5cm}
\caption{(Color online) Transverse momentum dependence of isotropic coefficient $v_0$ obtained from the azimuthal distributions of total radial velocity, multiplicity and mean radial velocity in Au+Au collision at $E_{\rm lab} = 10A$ and $40A$ GeV.}
\label{v0_pt}
\end{figure}
Comparing Figure\,\ref{v2-pt}(a) with Figure\,\ref{v2-pt}(b), we see a very weak (almost insignificant) energy dependence of $v_2$ in terms of all three variables concerned. We may reckon that FAIR-CBM energy range may not provide us with a proper platform to study the energy dependence of anisotropy, but it may be suitable to study the issues related to the isotropy measure $v_0$. The $p_{_T}$ dependence of $v_0$ has been shown in Figure\,\ref{v0_pt}. It is observed that the $v_0$ coefficients associated with $\langle N_{\rm ch} \rangle $, $\langle V_{_T}\rangle $ and $\langle \langle v_{_T} \rangle \rangle$ while plotted against $p_{_T}$, exhibit similar nature. In the low $p_{_T}$ region, the $v_0$ values extracted from each variable rise with increasing $p_{_T}$, attain a maximum and then fall off to a very small saturation value (almost zero) at both incident energies beyond $p_{_T}=1.25$ GeV/c. Once again, while $v_0(\langle V_{_T} \rangle)$ values at $E_{\rm lab}=10A$ and $40A$ GeV are almost identical, the $v_0(\langle N_{\rm ch} \rangle)$ values at $E_{\rm lab}=40A$ GeV are higher in the low $p_{_T}$ region ($p_{_T}\lesssim 0.7$ GeV/c) than those at $10A$ GeV. On the contrary, the $v_0(\langle \langle v_{_T} \rangle \rangle)$ values obtained at $E_{\rm lab}=40A$ GeV are lower in the low $p_{_T}$ region ($p_{_T}\lesssim 0.5$ GeV/c) than those at $10A$ GeV. At FAIR energy the random thermal motion of particles perhaps dominates over their collective behavior, which at high $p_{_T}$ leads to a very small amount of azimuthally integrated magnitude of net flow. 
\section{Conclusion}
\label{conclusion}
In this paper we present some basic results on the elliptic and radial flow of charged hadrons. The study is based on the azimuthal distributions of total transverse velocity, mean transverse velocity and multiplicity of charged hadrons produced in Au+Au collisions at $E_{\rm lab} = 10A$ GeV and $40A$ GeV. We have used the AMPT model (string melting version) to generate the events. We observe that azimuthal asymmetries are indeed present in all three distributions. However, we also note that in our simulation results the azimuthal anisotropy of the final state particles is predominantly due to the asymmetry of particle multiplicity distribution, and only a small fraction of this asymmetry is due to kinematic reasons. The overall nature of the dependence of the elliptic anisotropy parameter on the centrality of the collision and transverse momentum of produced particles remains similar for the three variables considered in the present analysis. The elliptical flow parameter is highest in the mid-central collisions, and within the interval $0 \leqslant p_{_T} \leqslant 2.0$ GeV/c it is highest at the highest $p_{_T}$. From our simulated results in the FAIR energy range we find a very small energy dependence of the elliptical flow parameter. On the other hand, the azimuthally integrated magnitude of the radial flow is maximum for most central collisions and its values are high in the low $p_{_T}$ region. From this analysis we see that the contribution to $v_0$ from the asymmetry in multiplicity distribution and that coming from the asymmetry in kinematic variable $v_{_T}$, exhibit an opposite incident beam energy dependence. While the former is higher at higher $E_{\rm lab}$, the latter is higher at lower $E_{\rm lab}$. Our simulated results are consistent with those obtained from RHIC and LHC energies, and do not require any new dynamics to interpret. However, in future there is enough scope to appropriately model these results in terms of relevant thermodynamic and hydrodynamic parameters associated with the intermediate ``fireball'' produced in $AB$ collisions. 

\end{document}